\documentclass[12pt]{article}
\usepackage[a4paper,width=6.8in,height=8.8in]{geometry}
\usepackage{amsmath,amssymb}

\usepackage{amsmath,amssymb,mathtools,amsthm,
	textcomp,mathscinet}
\usepackage{amsfonts,graphicx,enumerate,subcaption,placeins}
\usepackage{longtable}
\usepackage{setspace}
\usepackage[utf8]{inputenc}
\usepackage[english]{babel}
\usepackage{booktabs,multirow}
\usepackage{blkarray}
\usepackage{lscape}
\usepackage[T1]{fontenc}
\usepackage{authblk,color}
\usepackage[english]{babel}
\usepackage{amsfonts, amstext, amsthm, booktabs, dcolumn}
\usepackage{graphicx}
\usepackage{multirow}
\usepackage{float}
\usepackage{caption}
\usepackage{centernot}
\usepackage{times}
\usepackage[nottoc]{tocbibind}
\usepackage[mathscr]{eucal}
\usepackage{xcolor}
\usepackage[colorlinks=true,linkcolor=blue, allcolors=blue]{hyperref}%

\definecolor{gr}{rgb}{0.7, 0.0, 0.15}

\numberwithin{equation}{section}
\begin{document}
\title{American Options Pricing under Heston Model via Curriculum Learning in Coupled PINNs}
\author[a]{{\Large Rohan}}
\author[b]{{\Large Siddanth Shetty}}
\author[*]{{\Large Amit N. Kumar}}
\affil[ ]{Department of Mathematical Sciences,}
\affil[ ]{Indian Institute of Technology (BHU) Varanasi,}
\affil[ ]{Varanasi (Uttar Pradesh) - 221 005, India.}
\affil[a]{Email: rohan17217@gmail.com}
\affil[b]{Email: soddushetty2003@gmail.com}
\affil[*]{Email: amit.kumar2703@gmail.com}

\date{}
\maketitle

\vspace{-1cm} 
\begin{abstract}
\noindent
In American options, the early exercise feature allows the option to be exercised at any time prior to expiration. However, this flexibility introduces a challenge: the pricing model must value the option while simultaneously determining an unknown, time-varying exercise boundary. The Heston model is one of the most popular ways to model real market behavior because it allows volatility to change over time. However, unlike European options, there is no closed-form solution for American options under the Heston model, so we have to use numerical methods. In this paper, we propose a novel approach to solving the stochastic Heston partial differential equation for American options, using coupled physics-informed neural networks (PINNs) to predict both the option price and the free boundary, while employing curriculum learning and adaptive resampling to stabilize model training. Our work builds on recent deep learning methods but introduces a more effective training strategy to address the limitations of these approaches. The numerical results demonstrate the effectiveness of the proposed learning framework, providing a robust and efficient alternative to pricing American options, enabling rapid inference and accurate estimation under stochastic volatility.
\end{abstract}

\noindent


\section{Introduction}
The last few decades have witnessed a monumental rise in the trading of options derived on stocks, foreign exchanges and commodities markets. Most options traded in the stock markets today, especially in US and Canada are American options, which are a special type of options that can be exercised at any time up to and including expiration, in contrast to their European counterparts that can only be exercised on their expiration date. This early exercise feature introduces a free-boundary problem, significantly complicating the problem of valuing American options. This has led to a growing interest in research on American option pricing making it one of the most challenging problems in this domain.

\noindent
The Black-Scholes model(Black and Scholes \cite{black1973}) which models asset prices as geometric Brownian motion under the assumption of constant volatility was the first major breakthrough in the field of option pricing, laying the foundation for most of the research to come. However, the restrictive assumptions of this model limited its ability to capture key empirical features of financial markets. Contrary to the assumptions of the model, the returns of stocks in real-world markets have often been observed to deviate from the log-normal assumption of the Black–Scholes framework and a negative correlation is observed between asset prices and instantaneous variance, a phenomenon known as leverage effect. These dynamics produce the well known volatility smile/smirk, suggesting that the implied volatility is not constant but rather changes with different strike prices and maturity. Subsequent research in solving the pricing problem for American options saw the development of new models such as those based on jump-diffusion, Lévy processes, and stochastic volatility, to better reflect observed market behavior and overcome the limitations observed in the earlier frameworks. One such model that was prominent among these was the Heston model (Heston \cite{heston1993}) that modelled variance as a mean-reverting Cox–Ingersoll–Ross process while still remaining analytically solvable for European options. However, the valuation of American options under stochastic volatility frameworks gets further complicated in such a setting, where pricing now becomes a free-boundary problem governed by a multi-dimensional partial differential equation (PDE) for which no closed-form solutions have been derived to date. This has compelled researchers to increasingly rely on numerical methods for approximating their solutions, which in turn can end up being computationally intensive, and have limited utility in production. To address this computational problem, in this paper, we propose a new approach utilising a coupled PINNs framework to solve the Heston PDE for American options. We train the coupled PINNs simultaneously to predict both the option price and the free boundary in line with the PDE and boundary conditions of the Hestons model. Moreover, to address the problems of early convergence and spectral bias, we employ a curriculum learning strategy with adaptive resampling of data points, thereby stabilizing the overall model training process.

\noindent 
Historically, the Heston PDE for American options has been addressed using three main classes of numerical techniques: tree-based methods, Monte Carlo simulations, and deterministic PDE solvers. Lattice and tree-based methods approximate asset and volatility dynamics through discrete jumps, with the number of computational nodes typically growing quadratically with the number of time steps. Although these approaches are robust enough to handle the early exercise problem of American options, they become inefficient when dealing with stochastic processes that are correlated and in high-dimensional cases involving multiple stocks. Monte Carlo (MC) simulation methods, by contrast, estimate option prices by averaging payoffs across a large number of simulated paths. In particular, approaches such as Least Squares Monte Carlo (LSMC) (Longstaff and Schwartz \cite{longstaff2001}) provide a robust framework for high-dimensional problems by employing regression techniques to approximate continuation values. Although Monte Carlo methods are powerful for solving complex problems, they are computationally expensive and inherently have slow convergence rates of order $O(n^{-1/2})$, which means we may need to sample over a huge number of paths possibly several hundred thousands or even millions to get a reliable price with low error.

\noindent
In contrast to stochastic approaches that depend on random sampling, classical deterministic numerical methods for PDEs, notably the finite difference method (FDM) and the finite element method (FEM), offer grid-based accuracy and provable convergence for problems of low to moderate dimensionality. For instance, Hanson and Yan \cite{hanson2007} and Balajewicz and Toivanen \cite{balajewicz2012} employed FDM to study models incorporating stochastic volatility and jump-diffusion dynamics, while Cheang {\em et al.} \cite{cheang2013} developed transform-based techniques for related problems. Significant progress in the pricing of American options has been achieved through alternating direction implicit (ADI) schemes, such as those proposed by Zhu and Chen \cite{zhu2011b}, which are particularly effective in handling mixed spatial-derivative terms arising from the correlation between asset price and variance. Additionally, specialized approaches like the fitted finite volume method (FVM), introduced by Koffi and Tambue \cite{koffi2019}, have been designed to ensure stability and second-order convergence in multi-asset frameworks. Although these methods can be very accurate and give reliable solutions for the American option pricing probem, they have two practical limitations. Firstly, these methods require the construction of a carefully designed mesh over the entire sample space which becomes challenging as high-dimensional terms and multiple boundary conditions are involved. Secondly, in the case of American options involving the free boundary that is generally non-smooth, grid-based methods tend to struggle producing oscillations, loss of accuracy and slow convergence near the free boundary points.

\noindent
Since classical numerical methods suffer from issues related to mesh construction as highlighted above, meshless approaches, specifically ones based on deep learning have recently seen adoption in the field of option pricing. Literature on deep learning in this domain of late can be broadly categorized into two main branches. The first of these branches treats pricing as a supervised learning problem, in which neural networks are trained offline to approximate the mapping from model parameters and contract features to option prices or implied volatilities. Although, the origin of such approaches can be traced back to the early nonparametric work of Hutchinson {\em et al.} \cite{hutchinson1994}, it has only become substantially more practical recently with the advent of modern deep architectures and large-scale optimization techniques. For example, Olorunnimbe and Viktor \cite{olorunnimbe2023} provide a systematic survey of deep learning in financial markets, highlighting how neural networks have become the technique of choice for tasks ranging from derivative pricing to portfolio management while emphasizing the critical role of backtesting for real-world practice. Similarly, Horvath {\em et al.} \cite{horvath2021} showed that the deep learning framework is effective not only for pricing but also for calibrating stochastic and rough volatility models. A key advantage of these data-driven approaches lies in their ability to enable extremely fast repeated evaluations once the network has been trained, making them particularly well suited for applications such as model calibration and real-time market-making.

\noindent
On the other hand, another branch of deep learning delves beyond using it merely as a regression tool and extends its utility to serve as a numerical solver for the underlying PDE or optimal stopping problem itself. In this direction, Han {\em et al.} \cite{han2018} enabled efficient approximations using neural networks by reformulating high-dimensional parabolic PDEs as backward stochastic differential equations(BSDEs). Similarly, Sirignano and Spiliopoulos \cite{sirignano2018} were among the first to introduce a mesh-free framework called the Deep Galerkin Method, in which neural networks are trained to satisfy the governing differential operator and associated constraints directly, without the need for mesh construction. Recent advances in scientific machine learning and optimization techniques have spurred the development of mesh-free neural approaches for option pricing. While Raissi {\em et al.} \cite{raissi2019}  formalized the general PINN framework, in which the PDE, along with its terminal and boundary conditions, is enforced through the loss function,  other early contributors, for example, Louskos \cite{louskos2021}, demonstrated that physics-informed techniques can be applied to option pricing problems. In the context of the Heston model, Hainaut and Casas \cite{hainaut2024} showed that PINN-based solvers can efficiently approximate option prices over a wide range of maturities and model parameters without requiring retraining for each new specification. For American-style options, Gatta {\em et al.} \cite{meshless2023} extended the PINN framework to free-boundary problems and proposed a simple yet effective training strategy to improve convergence in the Black–Scholes setting. These developments make PINNs a natural candidate for tackling the American Heston problem. However, the problem becomes tricky for American options, as the deep neural network must now learn not only the option price but also the evolving exercise boundary separating the continuation and exercise regions. In practice, a straightforward PINN formulation may suffer from instability, since the option price surface exhibits sharp gradients near the strike price and as maturity approaches. The free boundary also depends on both time and variance, requiring a more structured, carefully designed learning strategy beyond vanilla PINN training to achieve accurate, stable solutions.

\noindent
This paper presents several key contributions. In this paper, we propose a coupled PINN framework, tailored to the Heston American put problem, in which two distinct neural networks are trained simultaneously: a preliminary solution network that approximates the option price within the continuation region, and a secondary boundary network that learns the unknown optimal exercise surface. This approach is inspired by the early work from Gatta {\em et al.} \cite{meshless2023}, for solving the Black-Scholes PDE using neural networks. To enhance training stability, we adopt a three-phase curriculum learning strategy inspired by Krishnapriyan {\em al.} \cite{curriculum2021}. A progressive training strategy is developed to improve training stability and convergence compared to naive joint optimization.  The procedure begins with pretraining the solution network while keeping the boundary fixed, followed by a joint training phase in which both networks are optimized simultaneously. In the final stage, the terminal free-boundary condition is incorporated as a fine-tuning constraint. Furthermore, we employ an adaptive resampling scheme to periodically regenerate collocation points, with increased concentration near the evolving free boundary and close to maturity, regions where the solution is more complex and harder to approximate accurately. Once trained, the proposed framework enables fast and efficient inference, making it a compelling mesh-free alternative for repeated option valuation under stochastic volatility. 

\noindent
This paper is organized as follows. Section \ref{2} introduces the formulation of the Heston American put problem along with the proposed coupled PINN architecture. Section \ref{3} details the curriculum learning strategy, data generation process, and adaptive resampling procedure. Numerical results are presented in Section \ref{4}, followed by concluding remarks and directions for future research in Section \ref{5}.

\section{Methodology}\label{2}

\subsection{Mathematical Formulation of the Heston American Put}
The Black--Scholes model assumes that volatility remains constant over time; however, this assumption is not consistent with observed market behavior. To overcome this limitation, several stochastic volatility models, such as the Hull--White model and the Heston model, have been introduced, in which variance is treated as a stochastic process. Under the risk-neutral measure, the Heston model describes the joint evolution of the asset price $S_t$ and its variance $v_t$ through the following system of stochastic differential equations:
\begin{align}
    dS_t &= r S_t \, dt + \sqrt{v_t} S_t \, dW_t^S \label{eq: Hestoneq1} \tag{2.1} \\
    dv_t &= \kappa(\theta - v_t) \, dt + \sigma \sqrt{v_t} \, dW_t^v \label{eq: Hestoneq2} \tag{2.2} \\
    dW_t^S dW_t^v &= \rho \, dt. \label{eq: Hestoneq3} \tag{2.3}
\end{align}
Here, $r$ denotes the risk-free interest rate, $\kappa$ is the risk-neutral speed of mean reversion, $\theta$ represents the risk-neutral long-term average variance, $\sigma$ is the volatility of volatility, and $\rho$ captures the correlation between the two Brownian motions $dW_t^S$ and $dW_t^v$. The dynamics of $dS_t$ closely resemble those of the geometric Brownian motion underlying the Black--Scholes model. In contrast, the variance process $dv_t$ follows a Cox--Ingersoll--Ross (CIR) square-root process. Notably, the CIR specification ensures that the variance remains strictly positive, provided the Feller condition ($2\kappa\theta > \sigma^2$) holds.

\subsubsection{Heston PDE}
Let $P(S,v,t)$ denote the price of an American put option with strike price $K$ and maturity $T$. By constructing a self-financing portfolio and applying It\^o's lemma together with the no-arbitrage principle, we obtain a two-dimensional PDE governing the option price $P$, which is given by
\begin{equation}
    \frac{\partial P}{\partial t} + \frac{1}{2}vS^2\frac{\partial^2 P}{\partial S^2} + \rho\sigma vS\frac{\partial^2 P}{\partial S\partial v} + \frac{1}{2}\sigma^2 v\frac{\partial^2 P}{\partial v^2} + rS\frac{\partial P}{\partial S} + \kappa(\theta - v)\frac{\partial P}{\partial v} - rP = 0 .\label{eq:heston pde} \tag{2.4}
\end{equation}
American put options grant the holder the right to exercise at any time $t \in [0,T]$. The domain is therefore partitioned into two regions by a critical exercise boundary $S_f(v,t)$. In the continuation region, where $S > S_f(v,t)$, the option price satisfies the Heston PDE given in \eqref{eq:heston pde}. In contrast, in the exercise region where $S < S_f(v,t)$, it is optimal to exercise the option and receive the intrinsic payoff $K - S$. Hence,

\begin{equation}
P = K-S, \text{ when } S < S_f(v,t). \label{eq: heston_in_exercise_region} \tag{2.5}
\end{equation}

\subsubsection{Boundary and Terminal Condition}
At maturity, the option price coincides with its intrinsic value
\begin{equation}
    P(S,v,T) = \max(K - S, 0). \label{eq:terminal_condition} \tag{2.6}
\end{equation} 
To prevent arbitrage opportunities, the option price must always be greater than or equal to its intrinsic value, $\max(K - S, 0)$. Moreover, the option price is required to be continuous across the interface between the continuation and exercise regions. Consequently, at the optimal exercise boundary $S_f(v,t)$, the following condition must be satisfied.
\begin{equation}
    P(S_f(v,t), v, t) = K - S_f(v,t). \label{eq:value_matching} \tag{2.7}
\end{equation}
The condition \eqref{eq:value_matching} is referred to as the value matching condition. Furthermore, the option's delta must coincide with the slope of the payoff, which is $-1$, at the free boundary. Hence,
\begin{equation}
    \left. \frac{\partial P}{\partial S} \right|_{S=S_f(v,t)} = -1. \label{eq:smooth_pasting} \tag{2.8}
\end{equation}
The condition \eqref{eq:smooth_pasting} is referred to as the smooth pasting condition. Finally, at the upper spatial boundary \( S_{\max} \), the far-field boundary condition must be satisfied.
\begin{equation}
    P(S_{\max}, v, t) = 0. \label{eq:far_field} \tag{2.9}
\end{equation}
Although the existing literature may employ additional boundary conditions to accurately model the option price and the free boundary, our approach to these will be discussed in later sections. For now, the expressions \eqref{eq:heston pde} through \eqref{eq:far_field} together define the Heston PDE along with its associated terminal and boundary conditions.

\subsection{Background on PINNs}
PINNs were first introduced by Raissi {\em et al.} \cite{raissi2019} as a novel deep learning framework for solving PDEs. While conventional neural networks depend on learning the relationship between the input and output, PINNs take advantage of knowing the physics of the problem by integrating the relevant differential equations into the loss function. In the process of constructing a PINN, a neural network $\hat{U}(x,t;\theta)$, with $\theta$ being weights
and biases of the network, is defined to represent the actual solution $U$. The necessary derivatives are
then computed using automatic differentiation and used to evaluate the loss function.\\
The main loss function $\mathcal{L}(\theta)$ in a PINN model consists of three parts as follows
\begin{equation}
    \mathcal{L}(\theta) = \mathcal{L}_{PDE}(\theta) + \mathcal{L}_{BC}(\theta) + \mathcal{L}_{IC}(\theta), \label{eq:pinn_loss} \tag{2.10}
\end{equation}
where $\mathcal{L}_{\mathrm{PDE}}$ penalizes deviations from the governing PDE within the interior domain, while $\mathcal{L}_{\mathrm{BC}}$ and $\mathcal{L}_{\mathrm{IC}}$ penalize deviations from the boundary and initial conditions, respectively.\\
PINNs have demonstrated strong performance across a wide range of applications, including fluid dynamics, quantum mechanics, reaction--diffusion systems, and financial modeling. In a conventional PINN framework, the boundary is fixed, which facilitates the learning of boundary conditions. However, in the pricing of American options, the optimal exercise boundary is not known a priori and evolves continuously over time. Forcing a single network to simultaneously learn both the pricing dynamics and the moving boundary often leads to significant optimization instability and convergence issues. To address this challenge, we introduce our proposed architecture in Section \ref{2.3}. 

\subsection{Coupled PINN Architecture}\label{2.3}
\begin{figure}[htbp]
    \centering
    \includegraphics[width=1.0\textwidth]{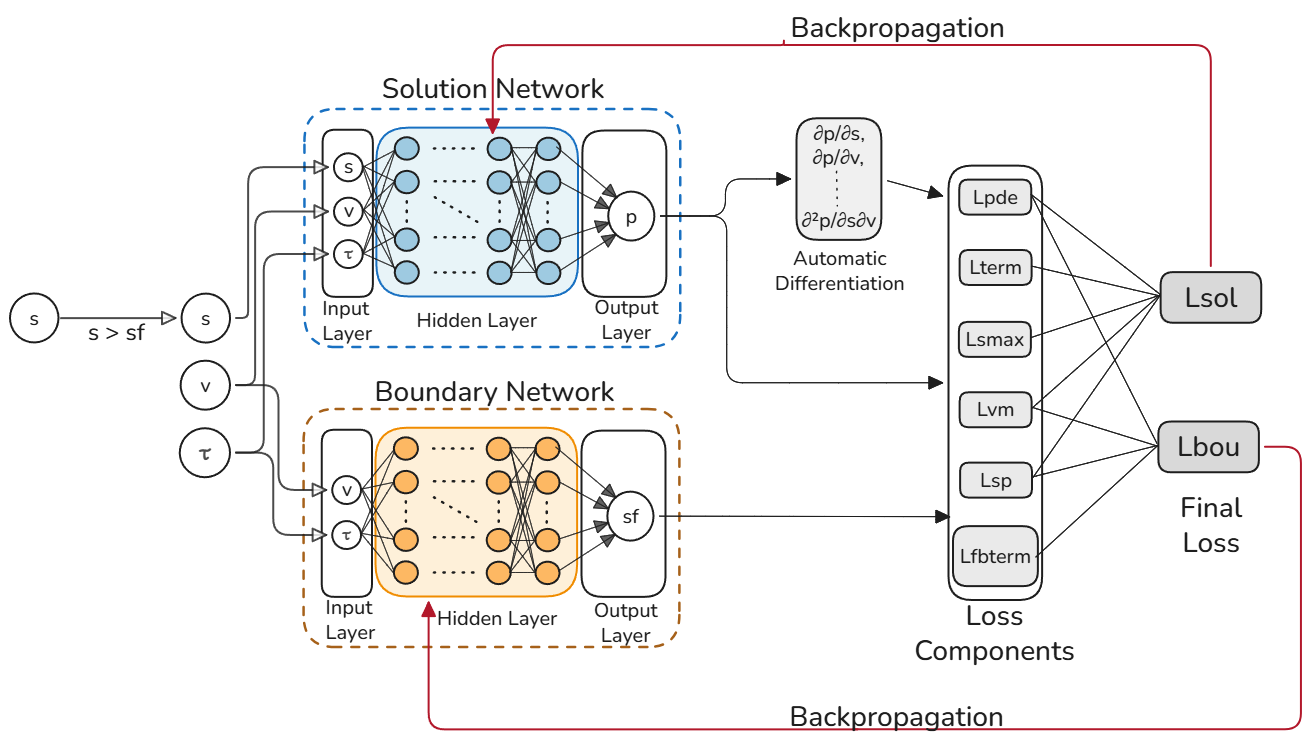}
    \caption{Overview of the proposed coupled PINN architecture.}
    \label{fig:architecture}
\end{figure}

In this paper, we employ a coupled PINN framework to solve the American put option pricing problem while simultaneously identifying the associated free boundary. Specifically, we train two distinct PINNs: a solution network, which approximates the option price in the continuation region, and a boundary network, which learns the unknown optimal exercise boundary that separates the continuation and stopping regions. The primary objective is to train these networks in a coupled manner, as the behavior of each network directly influences that of the other, resulting in a mutually dependent learning process.\\
The solution network approximates the normalized option price, with both inputs and outputs scaled by the strike price $K$ and the time to maturity $T$. Specifically, we define the normalized asset price as $s = S/K$ and the normalized time as $\tau = t/T$. Therefore,
\begin{equation}
    p(s, v, \tau; \theta) \approx \frac{P(S,v,t)}{K} \label{eq:norm_price} \tag{2.11}
\end{equation}
with the unscaled variance input $v$. Hence, the solution network aims to learn the mapping $N_{sol} : (s, v, \tau) \rightarrow p(\theta)$. Similarly, the boundary network approximates the normalized free boundary as
\begin{equation}
    s_{f}(v, \tau; \phi) \approx \frac{S_f(v, t)}{K}. \label{eq:norm_boundary} \tag{2.12}
\end{equation}
Hence, the boundary network aims to learn the mapping  $N_{bou} : (v, \tau) \rightarrow s_{f}(\phi)$.\\
To enhance stability and convergence in the coupled training framework, we adopt the following initialization strategy. The free-boundary network is initialized with a constant guess $s_{f}(\phi) \approx 0.9$ in normalized coordinates, rather than starting from zero. This provides a physically meaningful initial point, as the optimal exercise boundary for an American put option typically lies slightly below the strike price over most of the time horizon. In contrast, initializing at zero is unrealistic and can lead to instability during the early stages of training.\\
After initialization, the solution network is first trained to learn a reasonable approximation of the price surface under the current boundary estimate. Once the learned price surface becomes sufficiently reliable, we employ an alternating update scheme inspired by Gatta {\em et al.} \cite{meshless2023} for the Black--Scholes model. Rather than updating both networks simultaneously, we update the boundary network only once every 10 updates to the solution network. This strategy reduces the oscillatory behaviour and helps reduce the competition between the two networks.\\
In addition to the conditions \eqref{eq:terminal_condition}--\eqref{eq:far_field}, few more boundary conditions have been proposed in the literature. At $S = 0$, the underlying asset is assumed to have no chance of recovering. It means that the option holder should exercise immediately to receive the strike price $K$. Hence, some studies impose a Dirichlet boundary condition of the form $P(0, v, t) = K$. Alternatively, other works consider a discounted strike formulation, given by $P(0, v, t) = K e^{-r(T-t)}$. The existing literature presents differing perspectives on the appropriate boundary condition to impose when the variance approaches zero, that is, $v = 0$. Chockalingam and Muthuraman \cite{chockalingam2011} argue that the boundary behavior at $v = 0$ should be derived directly from the governing PDE. In contrast, Zhu and Chen \cite{zhu2011b} impose explicit Dirichlet conditions, namely $P(S,0,t) = \max(K - S, 0)$, along with a corresponding free boundary condition $S_f(0,t) = K$.  More recently, Nwankwo {\em et al.} \cite{nwankwo2025} implemented and compared both approaches to examine their respective impacts on pricing accuracy. However, there was no firm evidence that one condition should be preferred over the other, highlighting the need for further research into the boundary behavior as variance approaches zero.\\
In addition to the smooth pasting condition \eqref{eq:smooth_pasting}, several studies, including Ziogas and Chiarella \cite{ziogas2005} and Aitsahlia {\em et al.} \cite{aitsahlia2010}, impose an additional constraint at the free boundary. Since the payoff of a vanilla put, $\max(K - S, 0)$, is independent of the variance $v$, its derivative with respect to $v$ is identically zero. Consequently, at the boundary where the option price coincides with the payoff, the Vega of the option, $\frac{\partial P}{\partial v}$, must vanish. This yields the condition $\frac{\partial P}{\partial v}(S_f, v, t) = 0$.\\
In our formulation, we do not explicitly enforce the boundary condition at the lower bound of the asset price, nor do we impose the Vega condition at the free boundary. Given the absence of a clear theoretical advantage for any specific formulation, our framework deliberately avoids imposing an explicit, rigid boundary constraint as variance approaches zero. Artificially enforcing conditions such as $S_f(0,t) = K$ forces an unnatural, sharp kink in the free boundary geometry, which fundamentally destabilizes the neural network's optimization process. We allow the network to infer the appropriate behavior directly from the governing PDE, even for small values of the variance. We use one more condition which clamps the free boundary to the strike price at the time of maturity with the conditions \eqref{eq:heston pde}--\eqref{eq:far_field}. This condition follows from the fact that, at maturity, an American option reduces to its European counterpart, and it is optimal to exercise only when the asset price is below the strike price. Hence,
\begin{equation}
S_f(v,T) = K \text{ or } s_f(v,1) = 1. \label{eq:term_free_boundary} \tag{2.13}
\end{equation}
Based on the governing PDE, along with the associated initial and boundary conditions, we define separate loss functions corresponding to each component. Let $\mathcal{X}_{\mathrm{pde}}$, $\mathcal{X}_{\mathrm{term}}$, $\mathcal{X}_{S_{\max}}$, $\mathcal{X}_{\mathrm{bou}}$, and $\mathcal{X}_{\mathrm{fbterm}}$ denote the sets of collocation points associated with the PDE residual, terminal condition, far-field boundary, moving free boundary, and free-boundary terminal condition, respectively. Let $\mathcal{N}_{sol}(s, v, \tau)$ and $\mathcal{N}_{bou}(v, \tau)$ be the option price and free boundary as predicted by the respective networks. For the PDE loss, let $R[\cdot]$ represent the residual operator corresponding to the normalized Heston PDE, obtained from \eqref{eq:heston pde} through the transformation to the normalized variables $s$ and $\tau$. The loss within the continuation region ($s > s_{f,\phi}$) is then defined as
\begin{equation} \label{eq:loss_pde}
\mathcal{L}_{pde} = \left\| \mathcal{R}[\mathcal{N}_{sol}](s, v, \tau) \right\|^2_{(s,v,\tau) \in \mathcal{X}_{pde}}, \tag{2.14}
\end{equation}
where $|| . ||$ denotes the mean square error norm, and
\begin{equation} \label{eq:heston_residual} \tag{2.15}
\begin{split}
\mathcal{R}[\mathcal{N}_{sol}](s, v, \tau) &= \frac{1}{T} \frac{\partial \mathcal{N}_{sol}}{\partial \tau} + \frac{1}{2} v s^2 \frac{\partial^2 \mathcal{N}_{sol}}{\partial s^2} + \rho \sigma v s \frac{\partial^2 \mathcal{N}_{sol}}{\partial s \partial v} \\
&\quad + \frac{1}{2} \sigma^2 v \frac{\partial^2 \mathcal{N}_{sol}}{\partial v^2} + r s \frac{\partial \mathcal{N}_{sol}}{\partial s} + \kappa(\theta - v) \frac{\partial \mathcal{N}_{sol}}{\partial v} - r \mathcal{N}_{sol}.
\end{split}
\end{equation}
The other loss terms are derived in a similar way as
\begin{align} 
\mathcal{L}_{term} &= \left\| \mathcal{N}_{sol}(s, v, 1) - \max(1 - s, 0) \right\|^2_{(s,v) \in \mathcal{X}_{term}},\label{eq:loss_term} \tag{2.16}\\
\mathcal{L}_{Smax} &= \left\| \mathcal{N}_{sol}(s_{max}, v, \tau) \right\|^2_{(v,\tau) \in \mathcal{X}_{Smax}}, \label{eq:loss_smax} \tag{2.17}\\
\mathcal{L}_{VM} &= \left\| \mathcal{N}_{sol}(\mathcal{N}_{bou}(v, \tau), v, \tau) - (1 - \mathcal{N}_{bou}(v, \tau)) \right\|^2_{(v,\tau) \in \mathcal{X}_{bou}},\label{eq:loss_vm} \tag{2.18}\\
\mathcal{L}_{SP} &= \left\| \frac{\partial \mathcal{N}_{sol}}{\partial s}(\mathcal{N}_{bou}(v, \tau), v, \tau) + 1 \right\|^2_{(v,\tau) \in \mathcal{X}_{bou}},\label{eq:loss_sp} \tag{2.19}\\
\mathcal{L}_{fbTerm} &= \left\| \mathcal{N}_{bou}(v, 1) - 1 \right\|^2_{v \in \mathcal{X}_{fbTerm}}.\label{eq:loss_fbterm} \tag{2.20}
\end{align}
The total loss is constituted as follows:
\begin{align} 
\mathcal{L}_{sol} &= \mathcal{L}_{pde} + \mathcal{L}_{term} + \mathcal{L}_{S_{max}} + \mathcal{L}_{VM} + \mathcal{L}_{SP},\label{eq:loss_sol} \tag{2.21}\\
\mathcal{L}_{bou}& = \mathcal{L}_{pde} + \mathcal{L}_{VM} + \mathcal{L}_{SP} + \mathcal{L}_{fbTerm}.\label{eq:loss_bou} \tag{2.22}
\end{align}
The primary objective for both networks is to minimize their respective loss functions using the backpropagation algorithm. The overall architecture of the proposed framework is summarized in the following paragraph and illustrated in Figure \ref{fig:architecture}.

\noindent
As illustrated in Figure \ref{fig:architecture}, our framework consists of two neural networks operating in tandem: a price network, $N_{\mathrm{sol}}$, to predict the option's value, and a boundary network, $N_{\mathrm{bou}}$, to find the optimal exercise boundary. The procedure begins with the normalized inputs corresponding to the asset price ($s$), variance ($v$), and time ($\tau$). Prior to being passed through the input layers, we use the boundary network's current value to filter the spatial points. It ensures that only the points lying in the continuation region are fed as inputs. Consequently, the inputs ($s, v, \tau$) are fed into the input layer of $N_{sol}$, and ($v,\tau$) is fed into the input layer of $N_{bou}$.\\
In the next stage, automatic differentiation is employed to compute the required gradients of the price output. These derivatives, together with the network outputs, are then used to evaluate the six individual loss components. Based on conditions \eqref{eq:loss_sol} and \eqref{eq:loss_bou}, these components are aggregated to obtain the total loss functions for the solution and boundary neural networks, respectively. Finally, to account for the sensitivity of the moving boundary and to ensure stable training of $N_{\mathrm{sol}}$, we train the two networks at different speeds. Specifically, for every ten updates of the price network via backpropagation, the boundary network is updated only once. The updated boundary is then fed back into the filtering step, thereby forming a continuous and dynamically coupled training loop.

\section{Model Training}\label{3}

\subsection{Curriculum Learning}
\begin{figure}[h!]
    \centering
    \includegraphics[width=1.0\textwidth]{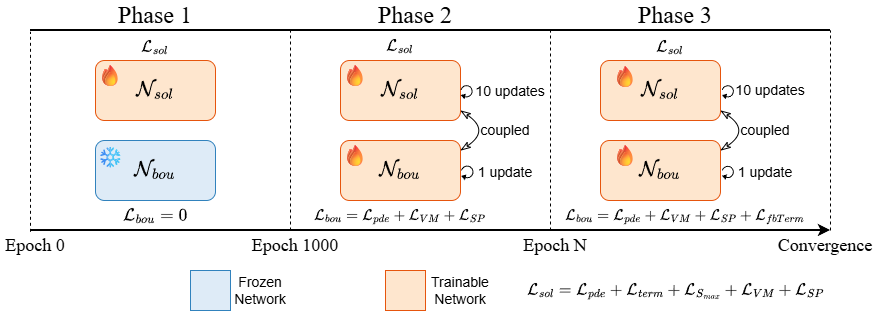}
    \caption{The three phases describing how training proceeds.}
    \label{fig:curriculum}
\end{figure}

Training the coupled neural network simultaneously on a moving free boundary is a challenging task. We observe that directly solving the full optimization problem, while simultaneously enforcing all constraints and boundary conditions, often leads to instability, causing the network to suffer from spectral bias. To avoid this instability and ensure convergence, we adopt a curriculum-based learning strategy. Instead of tackling the complete problem from the outset, we begin with an easier optimization task and gradually increase its complexity. Our architecture guides the model through three distinct phases, described below.\\
In the first phase, instead of training both networks simultaneously, we fix the boundary layer at a fixed starting value, $S_f = 0.9K$. Only the solution network is allowed to be trained for the first 1000 epochs. This allows $N_{sol}$ to learn the Heston PDE, along with the terminal and far-field boundary conditions, without the additional complexity introduced by a moving exercise boundary. Consequently, the value-matching and smooth-pasting losses quickly converge to fixed levels and remain constant after the initial few epochs, since $N_{bou}$ is not updated from its initial value.\\
After 1000 epochs of training, the solution network has learned the baseline physics of the Heston PDE. We then unfreeze the boundary network and allow the coupled training to proceed as described previously. However, in this second phase, we do not introduce all loss components simultaneously. In particular, we exclude the free-boundary terminal loss defined in \eqref{eq:loss_fbterm}. The primary reason for omitting this term is that it acts as a strong anchor, forcing the free boundary to output $K$ at maturity. This sudden kink creates large gradients, destabilizing training and preventing the other losses from converging. Instead of imposing this strict condition from the outset, we first allow the model to learn the overall shape of the boundary curve and the corresponding price dynamics in a more flexible manner. The network is trained under this configuration until 3500 epochs.\\
In the final phase, we also include the terminal boundary loss in the loss function. At this stage, the complete problem as defined in \eqref{eq:loss_sol} and \eqref{eq:loss_bou} and illustrated in Figure \ref{fig:architecture} is fully enforced. Since the networks have already learned the overall shape of the option price and the free boundary, this phase mainly serves as a fine-tuning step. Introducing the terminal constraint at this point allows the model to incorporate it smoothly without disrupting the global physics captured by the other loss components. We run this phase for 2000 epochs. This phase-wise learning strategy is illustrated in Figure~\ref{fig:curriculum}.

\subsection{Data Generation and Adaptive Resampling}
Instead of sampling data points directly in terms of $S$, $t$, and $v$, we sample using the corresponding dimensionless variables $s$, $\tau$, and $v$. For the dimensionless asset price $s = S/K$, we sample from 0 to 2. For dimensionless time to maturity $\tau = t/T$, we restrict the domain starting from 0 to 1. The variance values are already sampled from the range [0, 1], so they are not normalized. This normalization step helps neural networks better learn gradients and achieve stable, fast convergence.\\
In a standard PINN framework, collocation points are typically sampled according to a chosen distribution (for example, uniform, random, or Sobol) over the entire computational domain. However, as discussed earlier in \eqref{eq:heston pde} and \eqref{eq: heston_in_exercise_region}, the Heston PDE is valid only in the continuation region, that is, when the asset price satisfies $s > s_f$. If we were to sample points directly from the full domain $s \in [0,2]$ and enforce the PDE everywhere, the network would be forced to learn incorrect physics within the exercise region. To address this issue, we modify the data generation pipeline. We first generate collocation points for $\tau$ and $v$. Using them as inputs, we feed them into the partially trained boundary network $N_{\mathrm{bou}}$. Consequently, for any given pair $[\tau,v]$, we determine the approximate boundary location $s_f$ from the network and subsequently sample the asset price $s$ uniformly over the interval $[s_f,2]$.\\
Further, we identify two critical regions that require a higher density of training points. These regions typically correspond to boundaries, sharp curves/kinks in the graph's shape, or the steepest gradients. Interestingly, the two regions of interest come out to be: (i) near the free boundary in the continuation region, (ii) near the time of maturity. To ensure our model learns these regions well, we apply a mathematical power-law distribution to intentionally skew our input data towards them.

\subsubsection*{Clustering Near the Free Boundary}
Among the six loss components, learning the value-matching smooth-pasting condition is one of the most challenging, as it enforces a seamless transition between the option value and the intrinsic payoff at the free boundary. To capture this behavior accurately, we allocate a higher density of sample points in the boundary. At the start of training, when the boundary is initialized as $S_f = 0.9K$ or $s_f = 0.9$, we get the density distribution as illustrated in Figure~\eqref{fig:power_law_s}.

\subsubsection*{Clustering Near Time of Maturity}
\begin{figure}[htbp]
    \centering
    \begin{subfigure}[b]{0.45\textwidth}
        \centering
        \includegraphics[width=\textwidth]{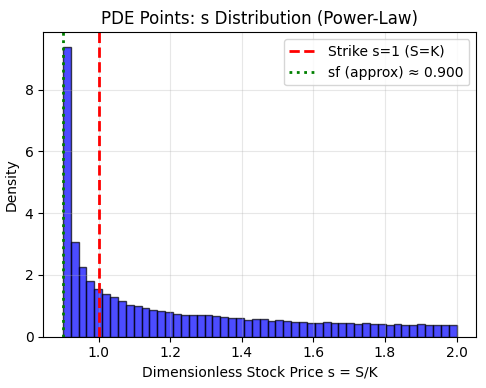}
        \caption{Data distribution for $s$}
        \label{fig:power_law_s}
    \end{subfigure}
    \hfill 
    \begin{subfigure}[b]{0.48\textwidth}
        \centering
        \includegraphics[width=\textwidth]{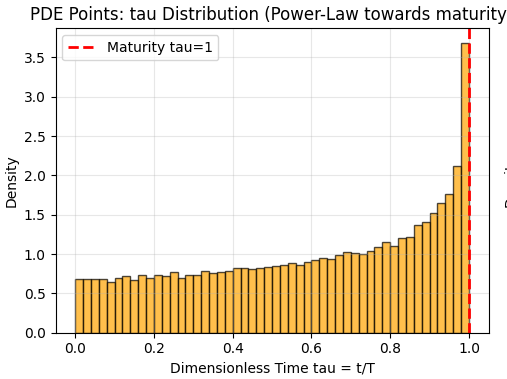}
        \caption{Data distribution for $\tau$}
        \label{fig:power_law_t}
    \end{subfigure}
    
    \caption{Power-law distributions used for biased sampling.}
    \label{fig:both_power_laws}
\end{figure}

A similar challenge occurs when we approach the time to maturity. The option value equals its intrinsic value, and there is a sharp kink in both the price function and the free boundary. To capture this, we allocate more sample points near the time to maturity. We get the density distribution as illustrated in Figure~\eqref{fig:power_law_t}.

\begin{figure}[H]
    \centering
    \includegraphics[width=0.73\textwidth]{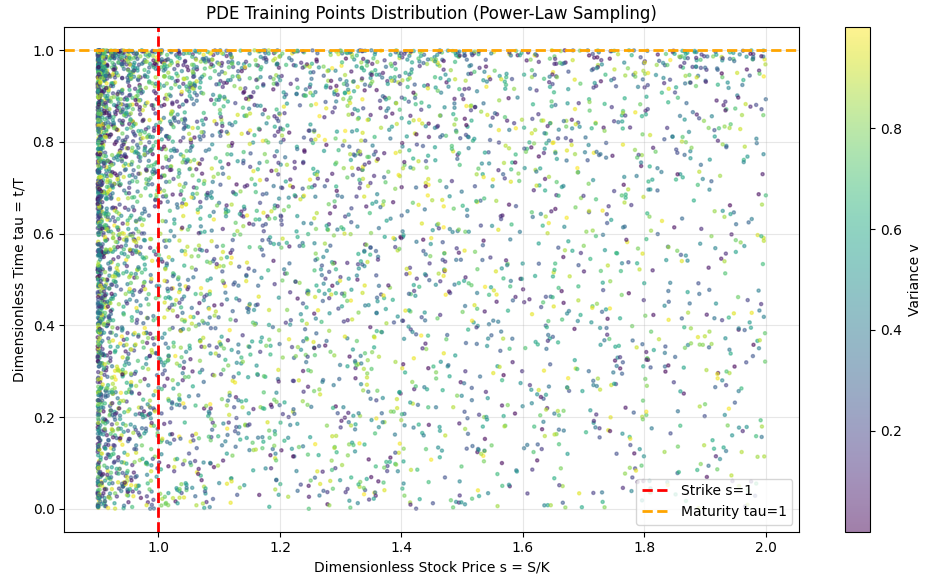}
    \caption{Scatter plot representing the distribution on a 2d grid of $s$ and $\tau$.}
    \label{fig:grid_sampling}
\end{figure}

\noindent
On a 2D grid, with the dimensionless asset price on the $x$-axis and the dimensionless time on the $y$-axis, we obtain the scatter plot illustrated in Figure~\ref{fig:grid_sampling}. The variance $v$ is sampled uniformly.

\noindent
Next, instead of generating our collocation points at the start of training and using them throughout, we completely resample all points at regular intervals (500-1000 epochs). It serves two main purposes. It first serves as a regularization technique, preventing overfitting to a specific set of data points. Secondly, it makes our data adaptive. As discussed earlier, we sample points closer to the free boundary; however, the free boundary is itself dynamic and constantly changing and refined by the coupled network. By periodically regenerating the data points based on the evolving free boundary, we ensure that a high concentration of data points follows the most important regions where our model may struggle to learn the optimal solution.

\subsection{Training}
The model consists of two separate connected neural networks, both employing the GELU (Gaussian Error Linear Unit) activation function (see, for details, Hendrycks and Gimpel \cite{hendrycks2023}). We prefer GELU over the standard ReLU (Rectified Linear Unit) activation because solving the Heston PDE involves second-order spatial derivatives, a twice-differentiable activation function is strictly required. The second derivative of ReLU is identically zero almost everywhere, which would cause the diffusion terms of the PDE to vanish during backpropagation. Note that
\begin{equation} \label{eq:relu} \tag{3.1}
\text{ReLU}(x) = \max(0, x)
\end{equation}
and
\begin{equation} \label{eq:gelu} \tag{3.2}
\text{GELU}(x) = x \Phi(x) = \frac{1}{2}x \left( 1 + \text{erf}\left(\frac{x}{\sqrt{2}}\right) \right),
\end{equation}
where $\Phi(x)$ is the cumulative distribution function of the standard normal distribution.

\begin{figure}[H]
    \centering
    \includegraphics[width=0.8\textwidth]{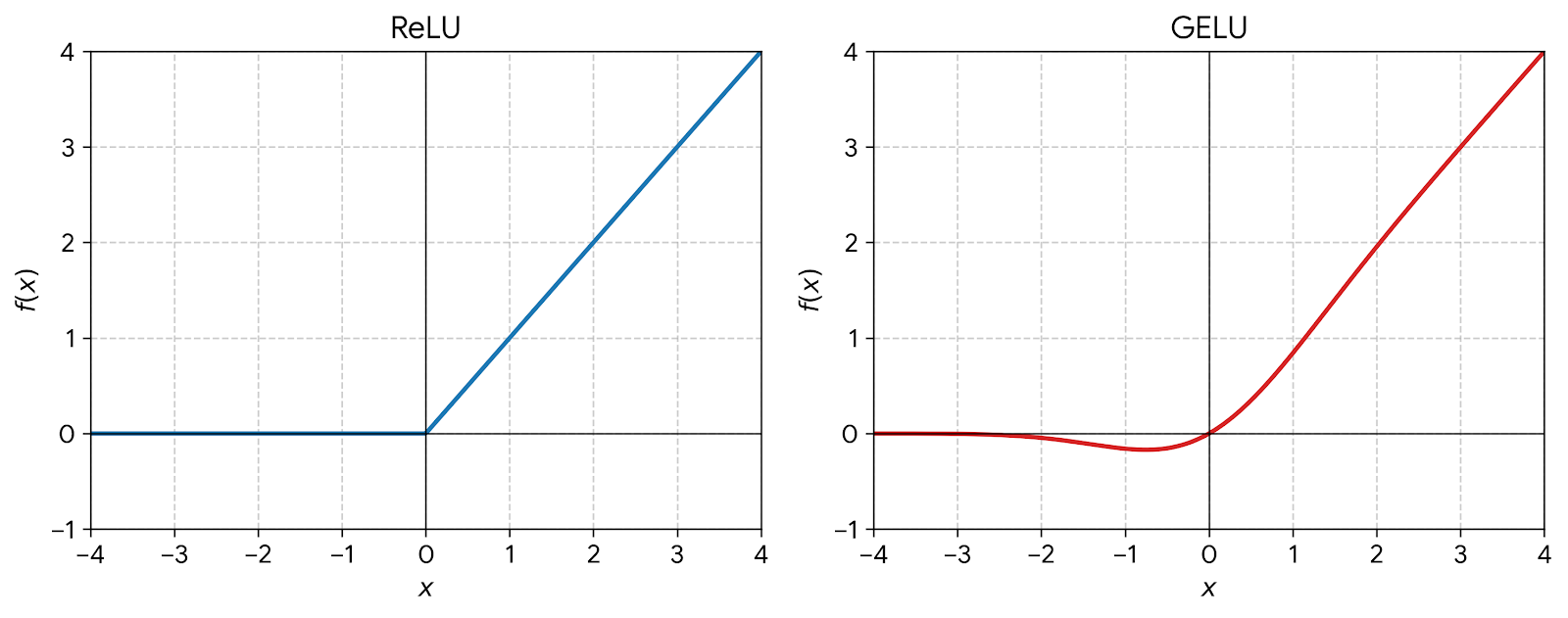}
    \caption{ReLU and GELU activation functions.}
\end{figure}

\noindent
The solution network consists of 5 hidden layers with the number of neurons in each layer given by $[32, 64, 128, 64, 32]$. The boundary network uses a different 5-layer structure: $[32, 64, 64, 64, 16]$. To ensure a stable start to training, we initialize the bias of the final layer of the boundary network to 0.9.\\
To implement the curriculum learning strategy, the training procedure is divided into two separate functions in our code. The first function manages Phases~1 and~2, while the second function is dedicated to Phase~3 of the training, as illustrated in Figure~\ref{fig:curriculum}. Since the networks face different loss-optimization challenges, we use different optimizers for each. The hyperparameters for the optimizers are given in Table~\ref{tab:hyperparameters}.

\begin{table}[htbp]
\centering
\begin{tabular}{|c|c|c|c|c|c|c|}
\hline
\multicolumn{3}{|c|}{} & Optimizer & LR & Step Size & Decay Factor \\
\hline
\multirow{2}{*}{Function 1} & \multirow{2}{*}{Phase 1 and 2} & $\mathcal{N}_{sol}$ & Adam & $10^{-3}$ & 500 & 0.9 \\ \cline{3-7}
 &  & $\mathcal{N}_{bou}$ & Adam & $10^{-3}$ & 500 & 0.9 \\
\hline
\multirow{2}{*}{Function 2} & \multirow{2}{*}{Phase 3} & $\mathcal{N}_{sol}$ & Adam & $5 \times 10^{-4}$ & 1000 & 0.9 \\ \cline{3-7}
 &  & $\mathcal{N}_{bou}$ & Adam & $5 \times 10^{-4}$ & 1000 & 0.9 \\
\hline
\end{tabular}
\caption{Hyperparameters for the coupled PINN architecture across training phases.}
\label{tab:hyperparameters}
\end{table}

\noindent
For each resampling cycle, a total of $150{,}000$ collocation points are generated. Among these, $\mathcal{X}_{\text{pde}} = 80{,}000$ points are sampled from the interior domain to enforce the PDE, while $\mathcal{X}_{\text{free}} = 40{,}000$ points are allocated to the free-boundary conditions (value matching and smooth pasting). The remaining points are distributed equally, with $10{,}000$ points each assigned to the terminal condition, terminal free-boundary condition, and spatial boundary conditions. During training, the entire dataset is shuffled and processed using a batch size of $8{,}192$.  Dividing the overall training process into multiple phases also enables the use of  different static loss weights for individual loss components. This flexibility allows the model to shift its focus toward specific losses that are more difficult to minimize, thereby improving overall training effectiveness.

\section{Results and Analysis} \label{4}
To evaluate the performance of our proposed coupled PINN architecture, we compare our results with well-established benchmarks proposed in the literature. The Heston model parameters used in our experiments are summarized in Table \ref{tab:heston_params}.

\begin{table}[htbp]
\centering
\setlength{\tabcolsep}{18pt} 
\renewcommand{\arraystretch}{1.4} 
\begin{tabular}{|c|c|c|c|c|c|c|c|}
\hline
$K$ & $T$ & $r$ & $q$ & $\kappa$ & $\theta$ & $\sigma$ & $\rho$ \\
\hline
10 & 0.25 & 0.1 & 0 & 5 & 0.16 & 0.9 & 0.1 \\
\hline
\end{tabular}
\caption{Heston model parameters used for benchmark evaluation.}
\label{tab:heston_params}
\end{table}

\noindent 
For this set of parameters, several studies, including Ikonen and Toivanen \cite{ikonen2009}, Clarke and Parrott \cite{clarke1999}, Zvan {\em et al.} \cite{zvan1998}, Thakoor {\em et al.} \cite{thakoor2018}, and Bolaky {\em et al.} \cite{bolaky2025}, have computed option prices for spot variances $v \in \{0.0625, 0.25\}$ and asset prices $S \in \{8, 9, 10, 11, 12\}$ at time $t = 0$, corresponding to three months until maturity. We present our results in comparison to these in Table \ref{tab:benchmark_comparison}. The last row of the table also reports the absolute error with respect to the reference values of Ikonen and Toivanen \cite{ikonen2009}.

\begin{table}[H]
\centering
\small 
\setlength{\tabcolsep}{10pt} 
\renewcommand{\arraystretch}{1.15} 
\resizebox{0.81\textwidth}{!}{  
\begin{tabular}{|c|l|c|c|c|c|c|}
\hline
\multirow{2}{*}{\textbf{Variance}} & \multirow{2}{*}{\textbf{{Reference}}} & \multicolumn{5}{c|}{\textbf{Asset Price ($S$)}} \\ \cline{3-7}
 & & \textbf{8} & \textbf{9} & \textbf{10} & \textbf{11} & \textbf{12} \\
\hline
\multirow{7}{*}{0.0625} 
 & Ikonen and Toivanen~\cite{ikonen2009} & 2.0001 & 1.1046 & 0.5129 & 0.2099 & 0.0820 \\
 & Clarke and Parrott~\cite{clarke1999} & 2.0000 & 1.1080 & 0.5316 & 0.2261 & 0.0907 \\
 & Zvan et al.~\cite{zvan1998} & 2.0000 & 1.1076 & 0.5202 & 0.2138 & 0.0821 \\
 & Thakoor et al.~\cite{thakoor2018} & 2.0000 & 1.1076 & 0.5200 & 0.2137 & 0.0820 \\
 & Bolaky et al.~\cite{bolaky2025} & 2.0001 & 1.1076 & 0.5203 & 0.2156 & 0.0830 \\
 & \textbf{Our Results (Proposed PINNs)} & \textbf{2.0000} & \textbf{1.1102} & \textbf{0.5187} & \textbf{0.2065} & \textbf{0.0778} \\ \cline{2-7}
 & \textit{Absolute Error} & \textit{0.0001} & \textit{0.0056} & \textit{0.0058} & \textit{0.0034} & \textit{0.0042} \\
\hline
\multirow{7}{*}{0.25} 
 & Ikonen and Toivanen~\cite{ikonen2009} & 2.0747 & 1.3257 & 0.7858 & 0.4401 & 0.2385 \\
 & Clarke and Parrott~\cite{clarke1999} & 2.0733 & 1.1329 & 0.7992 & 0.4536 & 0.2502 \\
 & Zvan et al.~\cite{zvan1998} & 2.0784 & 1.3337 & 0.7961 & 0.4483 & 0.2428 \\
 & Thakoor et al.~\cite{thakoor2018} & 2.0784 & 1.3336 & 0.7960 & 0.4483 & 0.2428 \\
 & Bolaky et al.~\cite{bolaky2025} & 2.0784 & 1.3336 & 0.7960 & 0.4487 & 0.2424 \\
 & \textbf{Our Results (Proposed PINNs)} & \textbf{2.0812} & \textbf{1.3393} & \textbf{0.7991} & \textbf{0.4473} & \textbf{0.2403} \\ \cline{2-7}
 & \textit{Absolute Error} & \textit{0.0065} & \textit{0.0136} & \textit{0.0133} & \textit{0.0072} & \textit{0.0018} \\
\hline
\end{tabular}}
\caption{Comparison of American put option prices under the Heston model at $t=0$.}
\label{tab:benchmark_comparison}
\end{table}

\begin{figure}[H]
    \centering
    \includegraphics[width=0.6\textwidth]{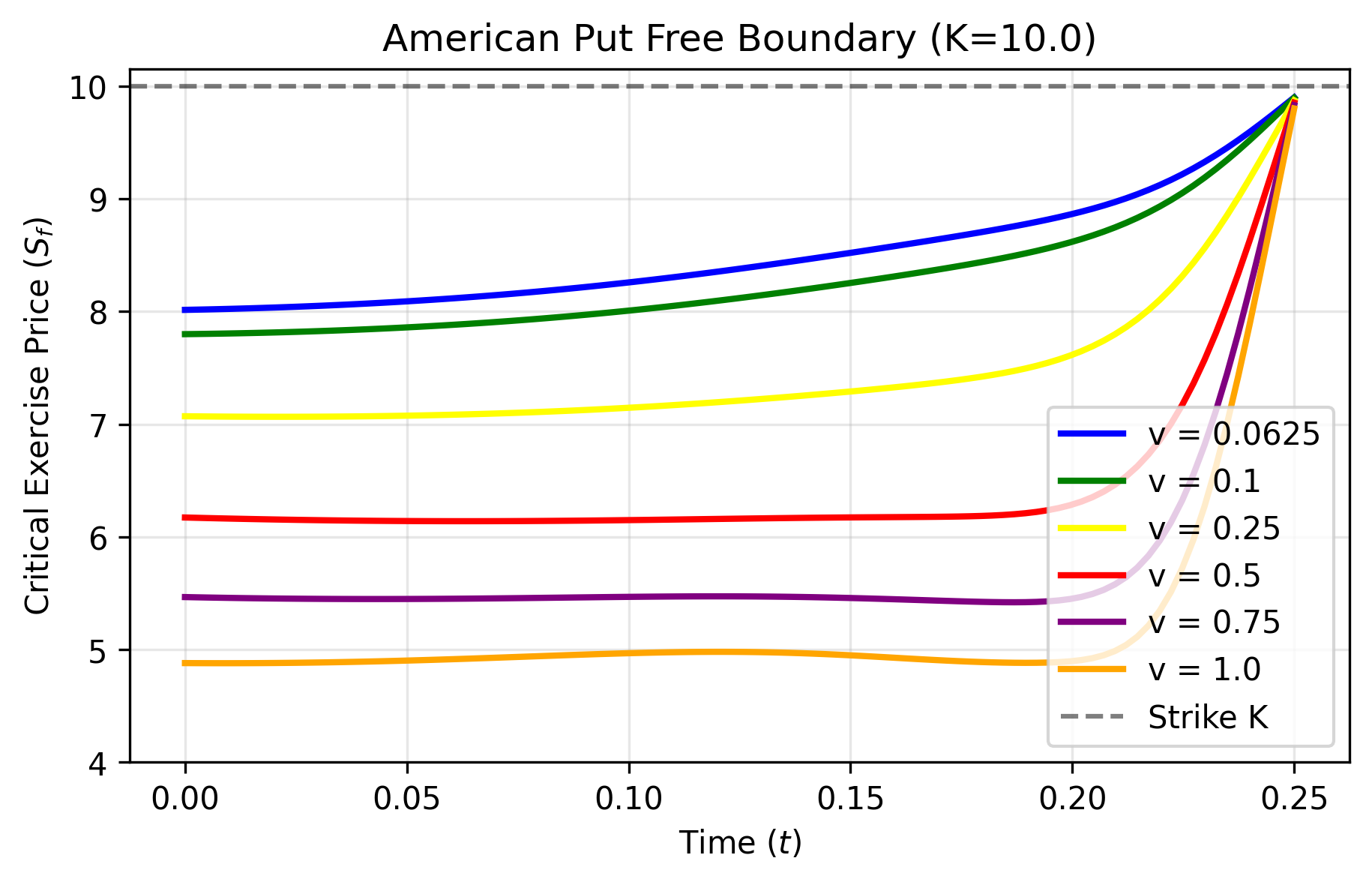}
    \caption{Evolution of the free boundary surface across time $t$ for various variance levels $v$.}
    \label{fig:free_boundary_surface}
\end{figure}

\noindent
Figure \ref{fig:free_boundary_surface} illustrates the free-boundary surface for different variance values plotted against time $t$. Since the variance represents the level of uncertainty in the underlying asset, the time value of the option increases with the variance level. Consequently, as the variance increases, the asset price that triggers the early-exercise condition decreases, which is evident from Figure~\ref{fig:free_boundary_surface}. Consistent with Table~\ref{tab:benchmark_comparison}, for $v = 0.0625$, the asset price $S = 8$ lies in the exercise region, and hence our model enforces the price to be $K - S$. In contrast, for $v = 0.25$, the same asset price falls in the continuation region, and the trained model predicts a value greater than 2.

\begin{figure}[H]
    \centering
    \includegraphics[width=0.8\textwidth]{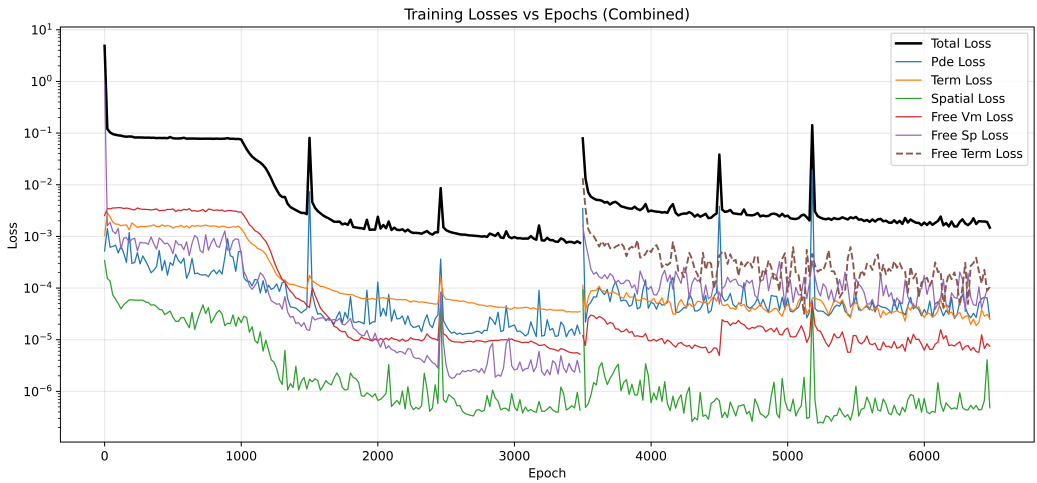}
    \caption{Training loss convergence over all epochs.}
    \label{fig:loss_over_training}
\end{figure}

\noindent
To understand the learning behavior and stability of the proposed model, we plot the loss convergence over the entire training period. Figure \ref{fig:loss_over_training} displays the evolution of the individual loss components along with the total loss, defined as the weighted sum of all individual losses in \eqref{eq:total_loss}.

\noindent
Note that
\begin{equation} \label{eq:total_loss}
\mathcal{L}_{total} = w_{pde}\mathcal{L}_{pde} + \dots + w_{fbTerm}\mathcal{L}_{fbTerm}.
\end{equation}

\noindent
The loss plot reflects the training strategy described in the methodology section. The three training phases are distinctly visible, along with the effect of dynamic resampling. During Phase 1 (0--1000 epochs), the boundary network $\mathcal{N}_{\mathrm{bou}}$ is frozen; consequently, after an initial sharp decrease in the loss during the first few epochs, the value-matching and smooth-pasting losses remain nearly constant. This phase allows the solution network $\mathcal{N}_{\mathrm{sol}}$ to stabilize before both networks are trained simultaneously. In Phase 2, the value-matching and smooth-pasting losses begin to decrease as the boundary network is updated. Owing to the coupled nature of the two networks, the PDE loss also decreases during this stage. At the beginning of Phase 3, when the final loss component is introduced, the networks require a few epochs to readjust before converging again. The few loss spikes in the graph are explained by the dynamic resampling of data points, which ensures the data distribution remains optimal throughout. 

\begin{figure}[H]
    \centering
    \includegraphics[width=1\textwidth]{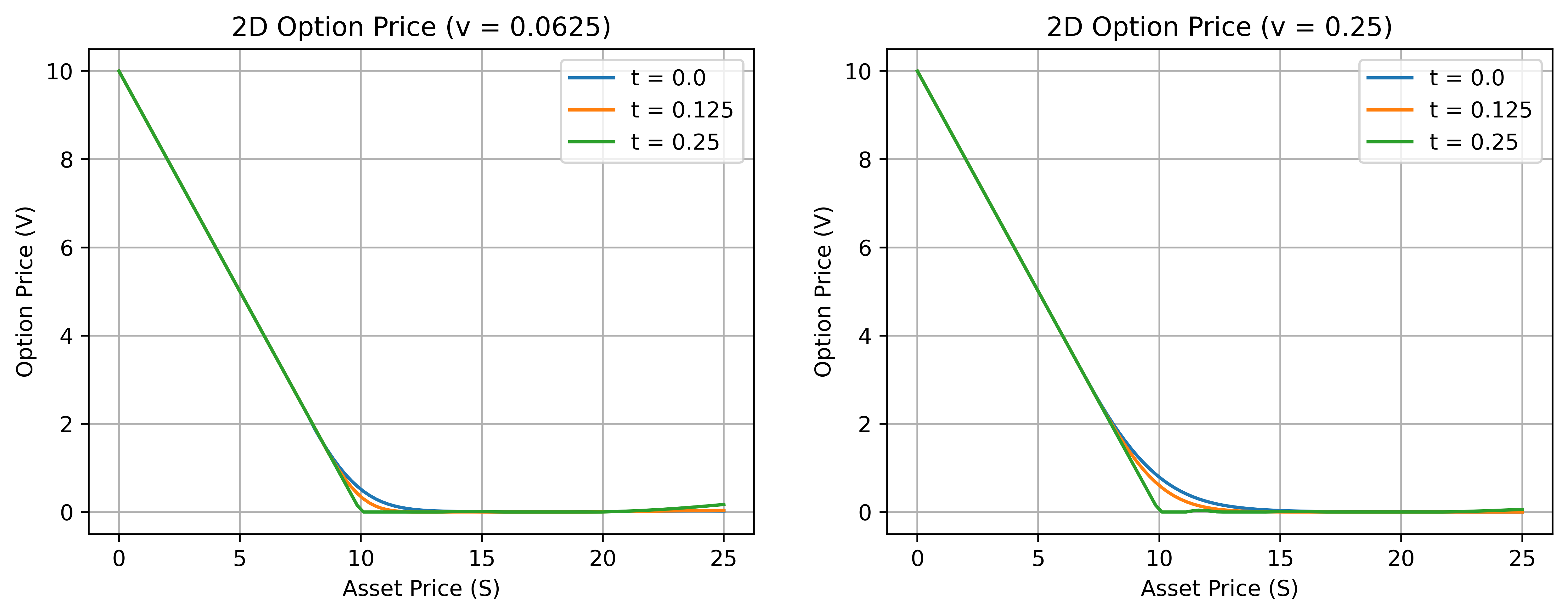}
    \caption{2D option price curves for variances $v=0.0625$ and $v=0.25$ at various times $t$.}
    \label{fig:price_vs_asset_1}
\end{figure}

\begin{figure}[H]
    \centering
    \includegraphics[width=0.8\textwidth]{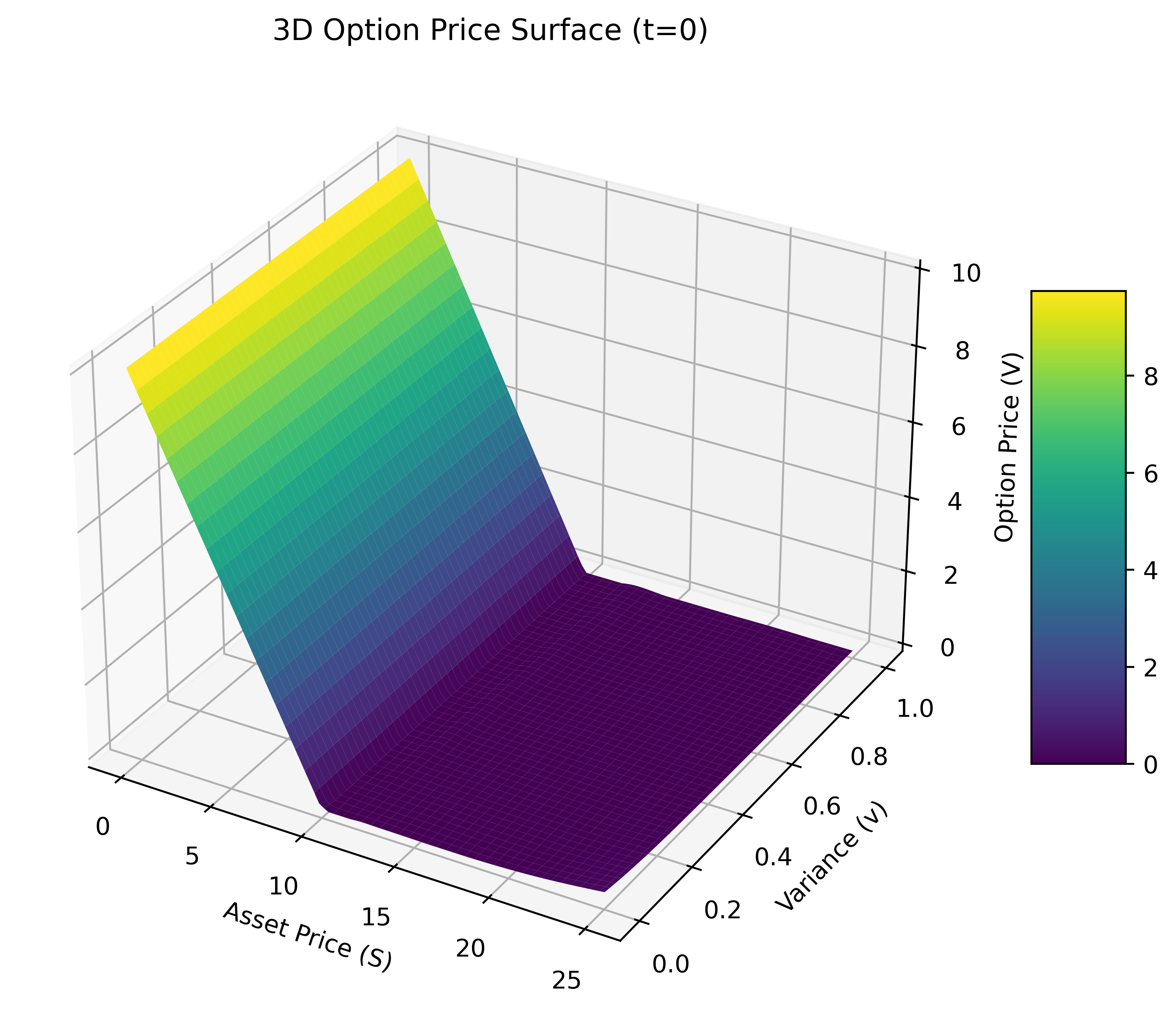}
    \caption{3D surface plot of the option price with respect to asset price and variance at time of maturity.}
    \label{fig:price_surface_3d}
\end{figure}

\noindent
Figure \ref{fig:price_vs_asset_1} illustrates the option price ($y$-axis) as a function of the asset price ($x$-axis) for different remaining times to maturity. As evident from the graphs, at maturity, the option price follows the payoff function $\max(K-S, 0)$. As the time to maturity increases, the option price for a given asset price also increases, reflecting the time value of the option.  Figure \ref{fig:price_surface_3d} presents a three-dimensional view of the option price at time of maturity, with the asset price and variance shown on the $x-$ and $y-$axes, respectively.


\noindent
To ensure the network properly captures the option's risk sensitivities, we compare the calculated Delta and Vega values with a standard MATLAB implementation of a finite difference method and the numerical results from Rouah \cite{rouah2013}. The Delta calculation is a direct computation of the asset price sensitivity by the network via automatic differentiation. However, since the PINN is defined using variance $v$ as a coordinate, the network's derivative actually computes the sensitivity with respect to $ v$. To compare this against the standard Vega, which is the derivative of the option value with respect to volatility $\sigma$, we must apply the chain rule. This computation is shown in the below expression

\begin{equation} \label{eq:vega_chain_rule}
\text{Vega} = \frac{\partial U}{\partial \sigma} = \frac{\partial U}{\partial v} \frac{\partial v}{\partial \sigma} = 2\sigma \frac{\partial U}{\partial v}.
\end{equation}

\begin{figure}[H]
    \centering
    \includegraphics[width=1\textwidth]{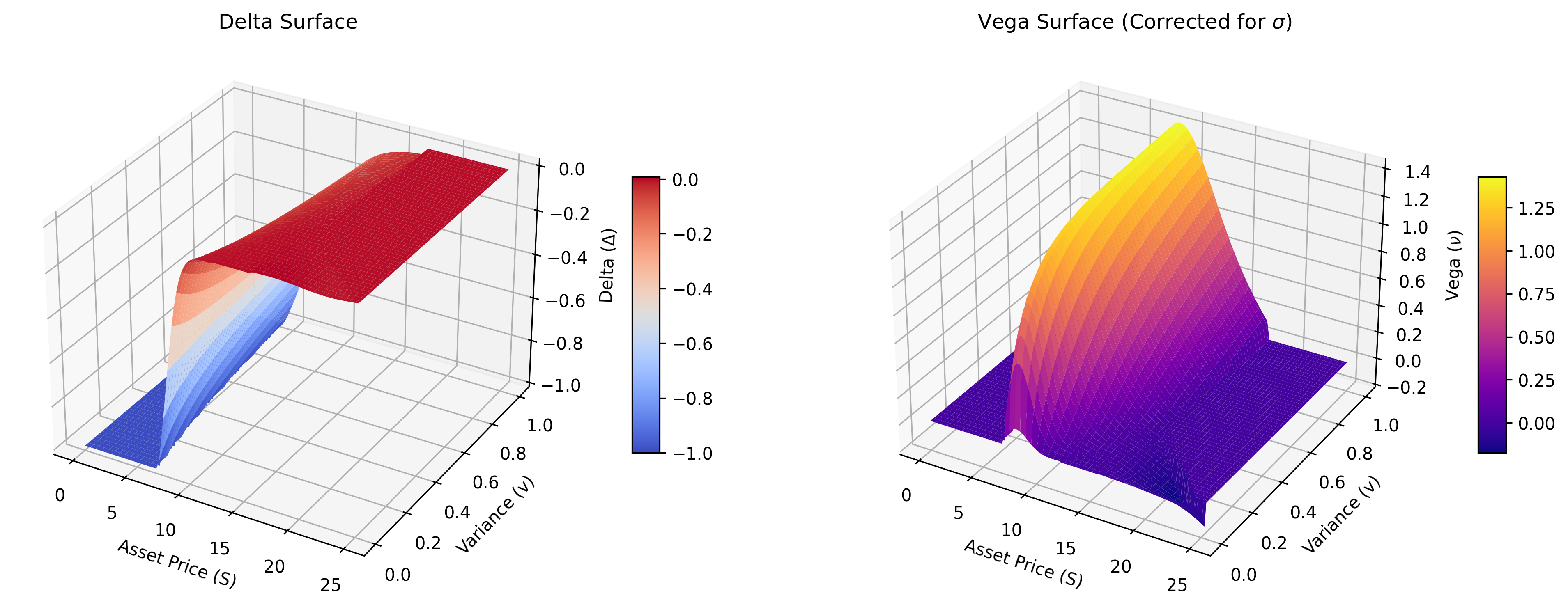}
    \caption{3D surface plots of the Delta and Vega risk sensitivities at $t=0$ and $v=0.0625$.}
    \label{fig:greeks_3d}
\end{figure}

\noindent
Table \ref{tab:greeks_comparison} provides a summary of the point-wise comparisons of the calculated values against a traditional implementation. The results provided by the coupled PINN framework, once adjusted for the chain rule, are very close to the traditional results. Figure~\ref{fig:greeks_3d} represents the 3D plots of the Delta and Vega surfaces.

\begin{table}[H]
\centering
\setlength{\tabcolsep}{10pt}
\begin{tabular}{|c|l|c|c|c|c|c|}
\hline
\multirow{2}{*}{\textbf{Greek}} & \multirow{2}{*}{\textbf{Reference}} & \multicolumn{5}{c|}{\textbf{Asset Price ($S$)}} \\ \cline{3-7}
 & & \textbf{8} & \textbf{9} & \textbf{10} & \textbf{11} & \textbf{12} \\ \hline
\multirow{3}{*}{Delta ($\Delta$)} 
 & Rouah~\cite{rouah2013} & -0.999 & -0.731 & -0.431 & -0.211 & -0.086 \\
 & MATLAB & -0.9474 & -0.7421 & -0.4476 & -0.2169 & -0.0892 \\
 & \textbf{Our Results} & \textbf{-1.0000} & \textbf{-0.7363} & \textbf{-0.4373} & \textbf{-0.2028} & \textbf{-0.0758} \\ \hline
\multirow{3}{*}{Vega ($\mathcal{V}$)} 
 & Rouah~\cite{rouah2013} & -0.005 & 0.848 & 0.983 & 0.675 & 0.317 \\
 & MATLAB & 0.0000 & 0.6907 & 0.8950 & 0.6909 & 0.3912 \\
 & \textbf{Our Results} & \textbf{0.0000} & \textbf{0.7170} & \textbf{0.8904} & \textbf{0.6902} & \textbf{0.3861} \\ \hline
\end{tabular}
\caption{Comparison of calculated Greeks (Delta and Vega) at $t=0$ and $v=0.0625$.}
\label{tab:greeks_comparison}
\end{table}

\begin{figure}[H]
    \centering
    \includegraphics[width=1\textwidth]{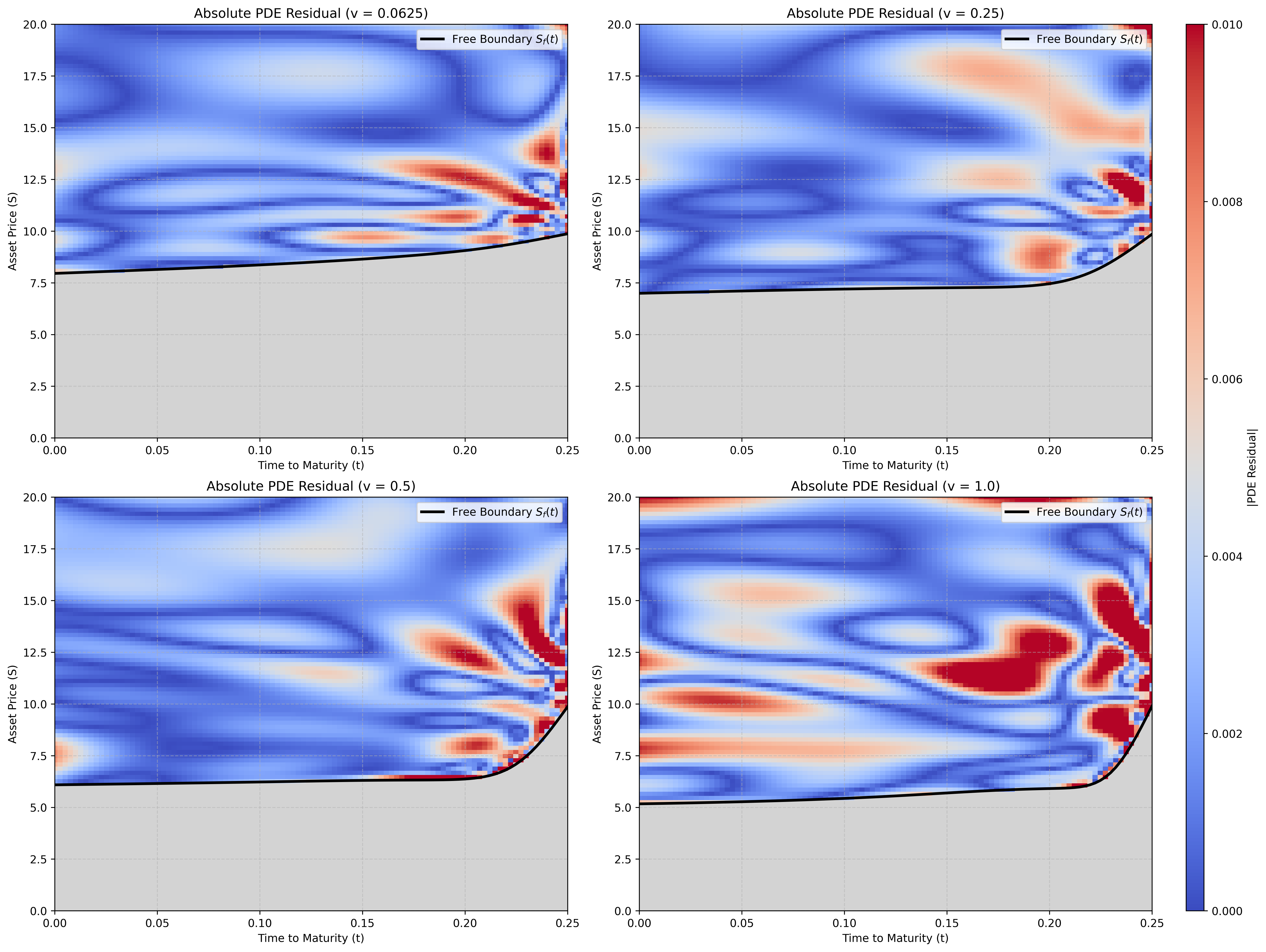}
    \caption{Absolute PDE residual heatmaps across varying volatilities utilizing adaptive resampling.}
    \label{fig:pde_residual_heatmap}
\end{figure}

\noindent
Next, Figure \ref{fig:pde_residual_heatmap} presents the heatmap of the absolute PDE residual in the continuation region for variance values $v \in \{0.0625, 0.25, 0.5, 1\}$. As expected, the model effectively minimizes the residual across most of the domain, particularly for lower variance levels. However, a few localized spikes persist near the free boundary and close to maturity. This behavior is not surprising from a mathematical standpoint, since these regions are highly nonlinear and thus particularly difficult for the network to approximate with precision.

\begin{figure}[H]
    \centering
    \includegraphics[width=1\textwidth]{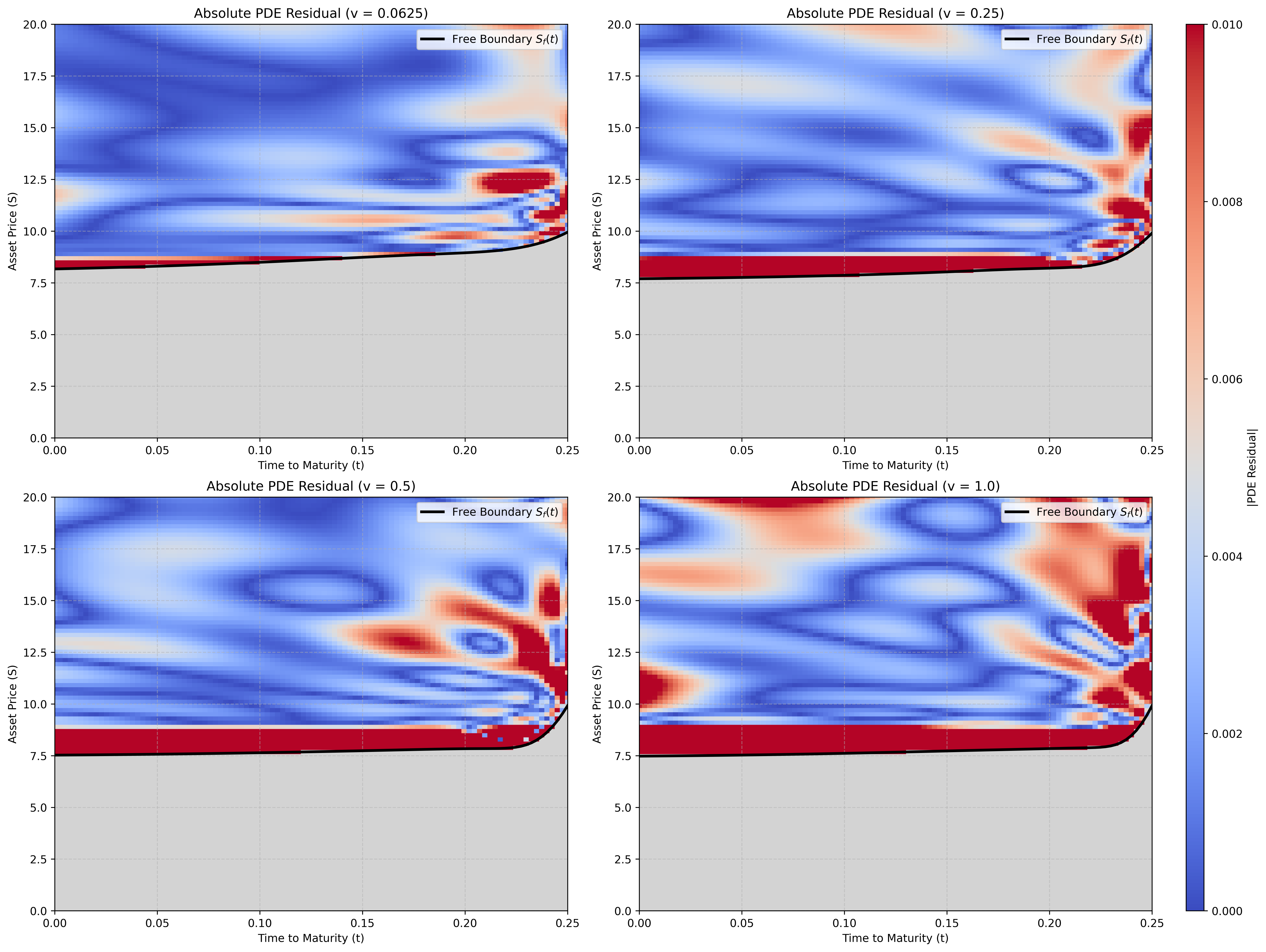}
    \caption{Absolute PDE residual heatmaps demonstrating high boundary errors without adaptive resampling.}
    \label{fig:high_residual_no_resample}
\end{figure}

\noindent
In addition, the observed error distribution highlights the role of the proposed resampling strategy. Before adaptive resampling was introduced, the model had difficulty capturing the pricing dynamics in these areas. This led to noticeable distortions in the free-boundary curve, especially at higher variance levels, along with increased PDE residuals near the boundary. As shown in the Figure~\ref{fig:high_residual_no_resample}, the residuals tend to peak in the region close to the free boundary. These effects are closely connected. When adaptive resampling is not used, the training process fails to concentrate enough collocation points in regions with strong nonlinearity. As a result, the network’s capacity to learn the correct solution behavior near the boundary remains limited.

\section{Conclusion and Future Scope}\label{5}
This paper proposes a novel mesh-free deep learning framework for determining the prices of American put options based on the Heston model. This problem is much more challenging than European options due to the early exercise boundary, which changes with time. Therefore, we propose an adaptive PINN network architecture for simultaneously capturing the option prices and the exercise boundary. Training both the networks simultaneously is not easy. The problem of directly optimizing the whole system results in instabilities and incorrect solutions. To address this issue, the approach of curriculum learning with a three-phase training schedule, which involves introducing the physical constraints step by step, has been suggested. In addition, the method uses adaptive resampling. More collocation points are assigned to regions that are harder to learn such as near the free boundary and as time approaches maturity. The computed option prices, the free boundary, and the Greeks are compared with standard benchmark. The result shows that the method achieves high accuracy across all these.

\noindent
Possible future directions for research include generalizing the method developed here to other types of options, for example, Asian and barrier options. It may also be possible to apply the method to stochastic volatility models beyond the Heston model, including the Bates model and models with rough volatility. Another direction to explore is applying the method to multi-dimensional variants of the Heston model. For such scenarios, the traditional mesh/grid methods become computationally costly and burdensome with increasing dimensions. On the contrary, mesh-free neural networks can handle such situations better, thus coupled PINNs may serve well in such scenarios.

\section*{Data Availability}
The source code used to generate the results in this study are available in a public GitHub repository at {\scriptsize \url{https://github.com/Rohan-217/American_Options_Pricing_using_Coupled_PINNs/tree/main}}.

\section*{Acknowledgements}
The third author is supported by MATRICS Research Grant (File No. MTR/2023/000200), India.


\bibliography{PA2PSDBib}

\end{document}